# Reinforcement Learning-Enabled Agent for Transmitter Optimization in Digital-Analog Radio-over-Fiber Fronthaul

Junhao Zhao, Huayuan Qin, Ouhan Huang, Zhongya Li, Chengxi Wang, Boyu Dong, Liangtao Chen, Xuyu Deng, An Yan, Penghao Luo, Renle Zheng, Yongzhu Hu, Aolong Sun, Yinjun Liu, Sizhe Xing, Nan Chi, and Junwen Zhang

***Abstract*—Digital-analog radio-over-fiber (DA-RoF) has emerged as a promising fronthaul solution that combines the high spectral efficiency of analog transmission with the robustness of digital transmission. However, the performance of DA-RoF critically depends on several tightly coupled parameters, including the rounding factor (RF), scaling factor (SF), geometric shaping (GS) factor, and pre-equalization taps coefficients, which jointly affect quantization noise, nonlinear distortion, and bandwidth-induced inter-symbol interference (ISI). Conventional grid search-based optimization is computationally prohibitive and impractical for optical communication. In this work, we propose a reinforcement-learning (RL)-enabled DA-RoF fronthaul agent architecture, capable of autonomously learning optimal transmitter parameters from end-to-end signal-to-noise ratio (SNR) feedback without a differentiable channel model. Experimental results demonstrate that the trained agent steadily improves SNR through sequential decision making and outperforms baseline, achieving ~2.7-dB SNR improvement for 1- to 4-order DA-RoF transmission, reaching final SNR of 35.8 dB, 42.9 dB, 53.8 dB, and 63.2 dB and supporting 1024-, 4096-, 16384-, 65536-quadrature amplitude modulation (QAM) format, respectively. These results validate that the proposed RL-enabled framework provides online, scalable, and hardware-efficient parameter optimization for DA-RoF fronthaul systems, paving the way toward high-order modulation format and intelligent next-generation radio access networks.**

## I. INTRODUCTION

As the research community pivots from 5th-generation mobile communication technology (5G) to 6th-generation mobile communication technology (6G), the cloud radio access networks (Cloud-RAN) architecture, which comprises a centralized unit (CU) for higher-layer protocol processing, a distributed unit (DU) for real-time baseband functions, and an radio unit (RU) that integrates the radio-frequency chain with massive-MIMO arrays, remains the main blueprint for RAN. However, 6G will push this split hierarchy to unprecedented extremes: sub-THz carrier frequencies, ultra-wide contiguous bandwidths, >1000-element intelligent surfaces, and ultra-high modulation formats are all under active discussion [1], [2], [3]. The mobile fronthaul (MFH) that links these segments must carry the raw wireless signals, posing stringent requirements on signal-to-noise ratio (SNR) and spectral efficiency (SE) [4], [5]. Radio-over-Fiber (RoF) technology is the most widely adopted interface solutions for MFH, as it determines the format in which wireless signals are transmitted over optical fibers. Depending on the requirements for SNR and SE, RoF has evolved into several distinct branches, including digital RoF (D-RoF), analog RoF (A-RoF), and digital-analog RoF (DA-RoF) architectures. For D-RoF, the Common Public Radio Interface (CPRI) has served as the standard interface for carrying fronthaul traffic in 4G networks and the early stages of 5G deployment. Although its large number of quantization bits guarantees high fidelity, it also translates into enormous bandwidth consumption. Specifically, a 1-GHz wireless channel requires 61.4 Gb/s over the fiber in a point-to-point configuration of long-term evolution (LTE) [6], [7]. With the rapid growth of mobile users, the limited spectral efficiency of CPRI has become insufficient to meet the ever-increasing bandwidth demands of data traffic. Therefore, the evolved CPRI (eCPRI) adopted in 5G alleviates this burden by compressing the payload and enabling different

Manuscript received xxx; revised xxx; accepted xx. This work is partially supported by Mobile Information Networks-National Science and Technology Major Project(2026ZD1308000), Natural Science Foundation of Shanghai (24ZR1490500), the Major Key Project PCL, and AI for Science Program, Shanghai Municipal Commission of Economy and Informatization (2025-GZL-RGZN-BTBX-02025) (Corresponding author: Junwen Zhang, email: junwenzhang@fudan.edu.cn).

J. Zhao, H. Qin, O. Huang, Z. LI, C. Wang, B. Dong, L. Chen, X. Deng, A. Yan, P. Luo, R. Zheng, Y. Hu, A. Sun, Y. Liu, S. Xing, N. Chi and J. Zhang are with the Key Laboratory for Information Science of Electromagnetic Waves (MoE), Shanghai Engineering Research Center of LEO Satellite Communication and Applications, Shanghai Collaborative Innovation Center of LEO Satellite Communication Technology, Future Information Innovative College (FIIC), Fudan University, Shanghai, 200433, China (email: jhzhao22@m.fudan.edu.cn; 24110720094@m.fudan.edu.cn; 23110720145@m.fudan.edu.cn; zhongyali20@fudan.edu.cn; 23210720253@m.fudan.edu.cn; boyudong@fudan.edu.cn; ltchen24@m.fudan.edu.cn; xydeng23@m.fudan.edu.cn; ayan22@m.fudan.edu.cn; phluo22@m.fudan.edu.cn; 24110720110@m.fudan.edu.cn; huyz23@m.fudan.edu.cn; alsun22@m.fudan.edu.cn; 23110720080@m.fudan.edu.cn; szxing21@m.fudan.edu.cn; nanchi@fudan.edu.cn; junwenzhang@fudan.edu.cn.

J. Zhao and H. Qin are also with the Shanghai Innovation Institute, Shanghai 200433, China.

Mentions of supplemental materials and animal/human rights statements can be included here.

Color versions of one or more of the figures in this article are available online at http://ieeexplore.ieee.org

functional splits [8]. However, D-RoF still exhibits relatively low SE. With the advancement of digital signal processing (DSP) technologies and the in-depth exploration of system architectures, extensive research efforts have also been devoted to A-RoF schemes with higher SE. A-RoF sends the wireless signals directly over the optical carrier, ideally occupying the same bandwidth as the wireless signals. In recent years, many studies have been devoted to this, such as intermediate-frequency over fiber (IFoF) [9], [10], [11], parallel intensity/phase (IM/PM) transmitters [12], and single-sideband (SSB) modulation with Kramers–Kronig detection [13]. They all offer very high SE and can reach terabit-per-second CPRI-equivalent capacities. The biggest disadvantage of these technologies, however, is sensitivity to transceiver non-linearity, which degrades SNR and limits modulation formats. With 1024-quadrature amplitude modulation (QAM) and beyond being targeted for future wireless systems [14], [15], conventional analog RoF falls short of the required fidelity [16]. To combine the fidelity of D-RoF with the SE of A-RoF, the DA-RoF technology has been proposed [17]. By applying simple rounding and differencing operations, the wireless signals are split into digital and analog waveform components, resulting in a substantial SNR improvement at the expense of linear spectral efficiency. DA-RoF technology has now been validated in several fronthaul architectures, including cascaded DA-RoF IM/DD links [18], NN-pre-equalized DA-RoF IM/DD links [19], lite-coherent systems with maximal-ratio combining [20], self-heterodyne coherent systems [21], [22], DA-RoF signal shaping for compensating nonlinear [23], and time-interleaved DA-RoF architecture [24]. These demonstrations collectively confirm the compatibility and practicality of DA-RoF as an emerging solution for future fronthaul deployments. However, the transmitter parameters in these works are usually selected by fixed configuration, empirical tuning, or grid search, while the coupled optimization among key DA-RoF parameters has not been sufficiently investigated. As a result, the obtained parameter settings may be suboptimal under practical hardware impairments and varying link conditions, limiting the achievable SNR improvement and deployment flexibility of DA-RoF systems. In DA-RoF systems, the rounding factor (RF), which governs the modulation order of the digital part, and the scaling factor (SF), which controls the amplitude of the analog part, significantly impact key performance metrics such as quantization noise, peak-to-average power ratio (PAPR) and SNR. Furthermore, considering the large-scale and high-density deployment characteristics of RUs in practical scenarios, DA-RoF fronthaul systems require the use of optical/electrical amplifiers at the CU/DU to boost signal power, thereby avoiding the need for additional amplifiers at the RUs that would otherwise increase deployment costs. However, excessively high transmit power can exacerbate signal nonlinearities, making geometric shaping (GS) of the signal a necessary technique. Similarly, in light of cost considerations at the receiver side, incorporating a pre-equalization module at the transmitter can effectively reduce the computational complexity of the DSP modules at receiver. When these design degrees of freedom are considered together with RF and SF of DA-RoF signals, they form a tightly coupled, joint-optimization problem. For instance, variations in the RF and SF can induce changes in PAPR, thereby influencing the average output power of the amplifier and subsequently the degree of signal nonlinearity, which lead to adjustments of the optimal GS parameters. Conversely, the adaptability of GS and pre-equalization to the channel can also impact the optimal selection of RF and SF parameters. Therefore, employing an exhaustive search approach to determine the parameters becomes computationally prohibitive and suboptimal, rendering it impractical for real-world deployment.

In the past, many efforts have been devoted to optimizing optical-communication systems. A popular line of work is end-to-end deep learning, where the transmitter, channel, and receiver are modeled as a differentiable pipeline and jointly optimized by gradient descent to reach a global optimum [25], [26], [27], [28], [29]. Although this approach is stable, it requires transmitter and receiver to host neural networks that perform modulation and demodulation. For densely deployed RUs, the resulting hardware and power overhead at the receiver side is often prohibitive. Furthermore, in a DA-ROF system, the transmitted waveform is produced by a simple, non-differentiable rounding operation, and the receiver does not need to introduce many extra operations to demodulate. Such an architecture is therefore fundamentally incompatible with end-to-end schemes that rely on a globally differentiable signal path for gradient-based optimization. More recently, reinforcement learning (RL) has emerged as an attractive alternative and has been widely adopted for large-scale model tuning [30] and embodied-intelligence tasks [31]. Unlike gradient-based end-to-end learning, RL allows an agent to interact with the environment modeled as a Markov decision process (MDP) and improve its policy through trial and error, thereby maximizing the expected long-term reward without requiring a fully differentiable channel model [32]. RL-based optimization has been investigated in communication systems for digital predistortion [33], [34], system optimization [35], and network resource allocation [36]. However, these methods mainly target general transmitter compensation, neural-network-based signal processing, or network-layer scheduling, and their state/action spaces are not designed for the DA-RoF signal decomposition mechanism. In DA-RoF transmission, RFs, SFs, geometric shaping (GS) factor, and pre-equalization coefficients are coupled through quantization noise, residual analog power, PAPR, nonlinear distortion, and bandwidth-induced ISI. The technical characteristics of RL align closely with the requirements of the DA-RoF fronthaul system. In this architecture, the intelligent agent can be deployed entirely at the CU/DU, using the signal quality feedback from the RUs as the reward to guide its training process. The RUs, whether currently deployed or yet to be deployed, do not require any additional modules or hardware modifications, since they simply provide feedback without executing the learning algorithm. This architecture is especially critical for cost-sensitive access networks, where minimizing complexity and cost at the remote RUs.

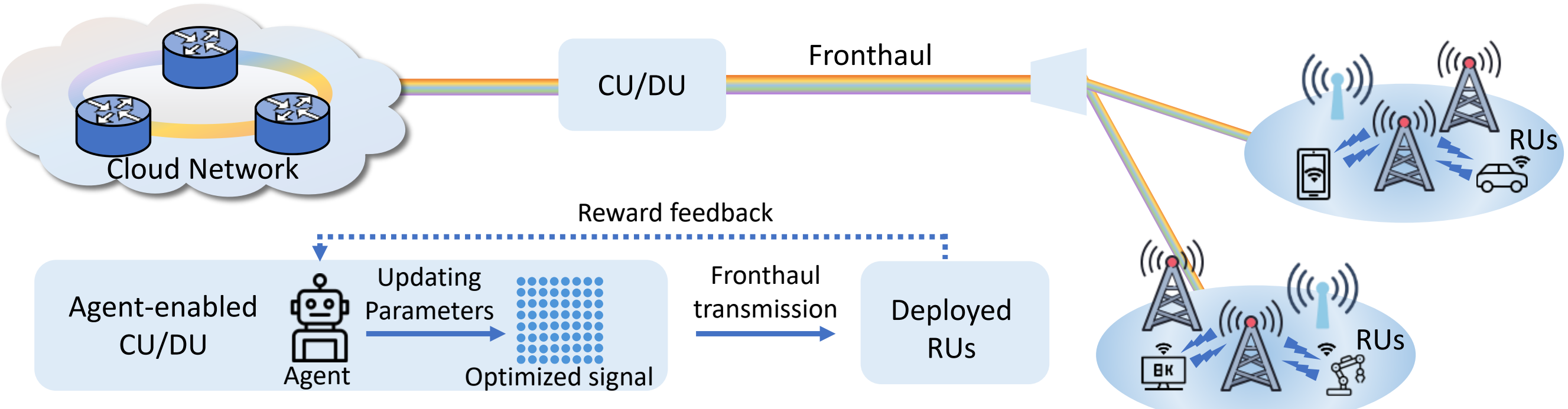

Fig. 1. RL-enabled DA-RoF fronthaul architecture.

In this work, we propose a reinforcement-learning (RL) enabled DA-RoF fronthaul agent for physical-layer performance optimization. A mathematical model of the DA-RoF intensity-modulation direct-detection (IMDD) system is developed, and the joint parameter optimization problem is formulated as an MDP, establishing the theoretical basis for RL-driven adaptive control. The agent is deployed entirely at the transmitter (CU/DU) and autonomously optimizes key DA-RoF parameters, including RFs, SFs, GS factor, and pre-equalization coefficients, without introducing any additional DSP cost on the receiver side. The agent is experimentally trained and validated, achieving 35.8-dB, 42.9-dB, 53.8-dB, and 63.2-dB SNR and supporting 1024-, 4096-, 16384-, and 65536-QAM formats for 1-, 2-, 3-, and 4-order DA-RoF signals, respectively. Compared with the grid-search baseline, an average SNR improvement of approximately 2.7 dB is achieved. These results confirm the effectiveness and practicality of the proposed framework in enhancing system performance with minimal complexity. The key contributions of this paper are summarized as follows:

- We formulate DA-RoF parameter optimization as an MDP and propose an RL-enabled fronthaul agent capable of model-free, closed-loop, and hardware-friendly optimization without relying on analytic channel models or exhaustive grid search.
- The proposed agent is deployed exclusively at the transmitter and jointly optimizes RF, SF, GS factor, and pre-equalization coefficients, maintaining low receiver complexity while ensuring compatibility with IMDD fronthaul deployment requirements.
- Experiments confirm that the proposed method achieves high-order QAM formats (up to 65536-QAM) and delivers ~2.7-dB SNR improvement over baseline for multiple DA-RoF orders, demonstrating strong performance gains, robustness, and deployment potential.

The remainder of the paper is organized as follows. Section II presents the mathematical modeling of the DA-RoF IMDD fronthaul system. Section III formulates the transmitter parameter optimization problem as an MDP. Section IV introduces the proposed RL-enabled DA-RoF fronthaul agent. Section V provides the experimental evaluation followed by the training and inference results of the proposed agent. Finally, the paper is concluded in Section VI.

## II. System Model

In this section, we establish the models of DA-RoF and system impairments to mathematically formulate the optimization problem of the DA-RoF fronthaul system, thereby providing theoretical guidance for optimization.

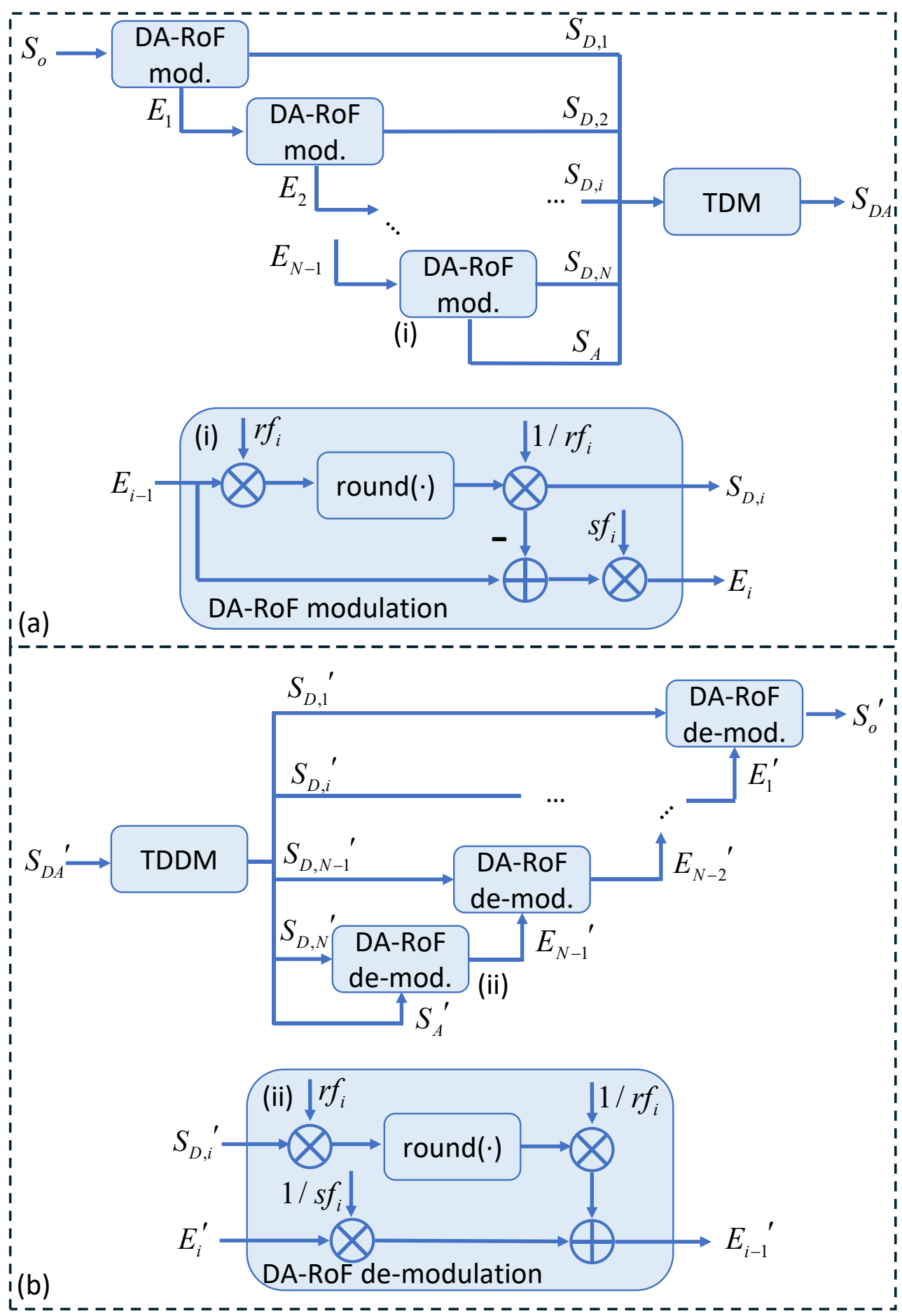

Fig. 2. Processing flow of N-order DA-RoF (a) modulation, and (b) de-modulation. Detailed operation of inset (i) DA-RoF modulation (DA mod.) and inset (ii) de-modulation (DA de-mod.) in the i-th DA-RoF. mod.: modulation; de-mod.: de-modulation; TDM: time division multiplexing; TDDM: time division de-multiplexing.

*A. DA-RoF transmission model*

Fig. 2 outlines the N-order DA-RoF modulation and demodulation pipeline. As illustrated in Fig. 2 (a), an N-order DA-RoF modulation splits the normalized wireless signal $S_o$ into digital parts $S_{D,i}$ and one analog part $S_A$ using DA-RoF modulation module. The signal processing of a single module is shown in inset (i) of Fig. 2, where the i-th digital part and residual analog error are expressed as

$$S_{D,i} = \begin{cases} round(rf_i \cdot S_o)/rf_i, \; i=1 \\ round(rf_i \cdot E_{i-1})/rf_i, \; i=2,3,...,N \end{cases} \quad (1)$$

$$E_i = \begin{cases} (S_o - S_{D,i}) \cdot sf_i, \; i=1 \\ (E_{i-1} - S_{D,i}) \cdot sf_i, \; i=2,3,...,N \end{cases} \quad (2)$$

$$S_A = E_N$$

where $rf_i$ and $sf_i$ are RF and SF of the i-th DA-RoF modulation. In the $i$-th modulation stage, the residual analog signal $E_i$ is scaled to the interval $[-rf_i, rf_i]$. A rounding operation then quantizes it, yielding a $(2 \cdot round(rf_i)+1)$-QAM digital part $S_{D,i}$. Subtracting $S_{D,i}$ from $E_{i-1}$ produces a new residual analog signal $E_i$. To strengthen the residual analog part, a gain $sf_i$ is applied to improve SNR and balance the power allocation of the DA-RoF signal $S_{DA}$, thereby controlling the overall PAPR. Therefore, the adaptation of parameters $rf_i$ and $sf_i$ to the actual channel environment directly affects the SNR of the recovered wireless signal. Furthermore, an N-order DA-RoF transmitter repeats procedure N times to obtain N digital parts and one residual analog part, which are finally time-division multiplexed for transmission. Each cascade stage typically delivers ~10 dB SNR improvement, at the cost of reduced spectral efficiency [18].

The processing flow of N-order DA-RoF de-modulation is as illustrated as Fig. 2 (b), whose detailed process is shown in inset (ii). The received DA-RoF signal $S_{DA}'$ is time division de-multiplexed to obtain the received digital parts and analog part $[S_{D,N}', S_{D,N-1}', ..., S_{D,1}', S_A']$. These signals are sequentially processed by DA-RoF de-modulation module to recover the original wireless signal $S_o'$. These operations can be mathematically expressed as follows:

$$S_o' = \sum_{i=1}^{N} [\frac{round(rf_i \cdot S_{D,i}')}{rf_i} \prod_{k=1}^{i-1} \frac{1}{sf_k}] + (\prod_{k=1}^{N} \frac{1}{sf_k}) S_A' \quad (3)$$

where $S_{D,i}'$ and $S_A'$ denote the received i-th digital part and analog part. In the following analysis, we consider an additive white Gaussian noise (AWGN) channel with a noise power of $P_N$. Since the DA-RoF signal is transmitted in a TDM manner, each sub-signal experiences the same level of white noise. In the DA-RoF de-modulation module, the noise of digital parts is eliminated by $round(\cdot)$ operation under error-free conditions. However, imperfections in the channel and hardware components can introduce signal impairments such as inter-symbol interference (ISI) and nonlinear distortion. Therefore, the symbol decision errors are introduced as

$$q_i = \frac{round(rf_i \cdot S_{D,i}')}{rf_i} - S_{D,i} \quad (4)$$

where $q_i$ is the symbol decision errors. Correspondingly, the symbol error rate (SER) can be expressed as

$$SER_i = \Pr[q_i \neq 0] \quad (5)$$

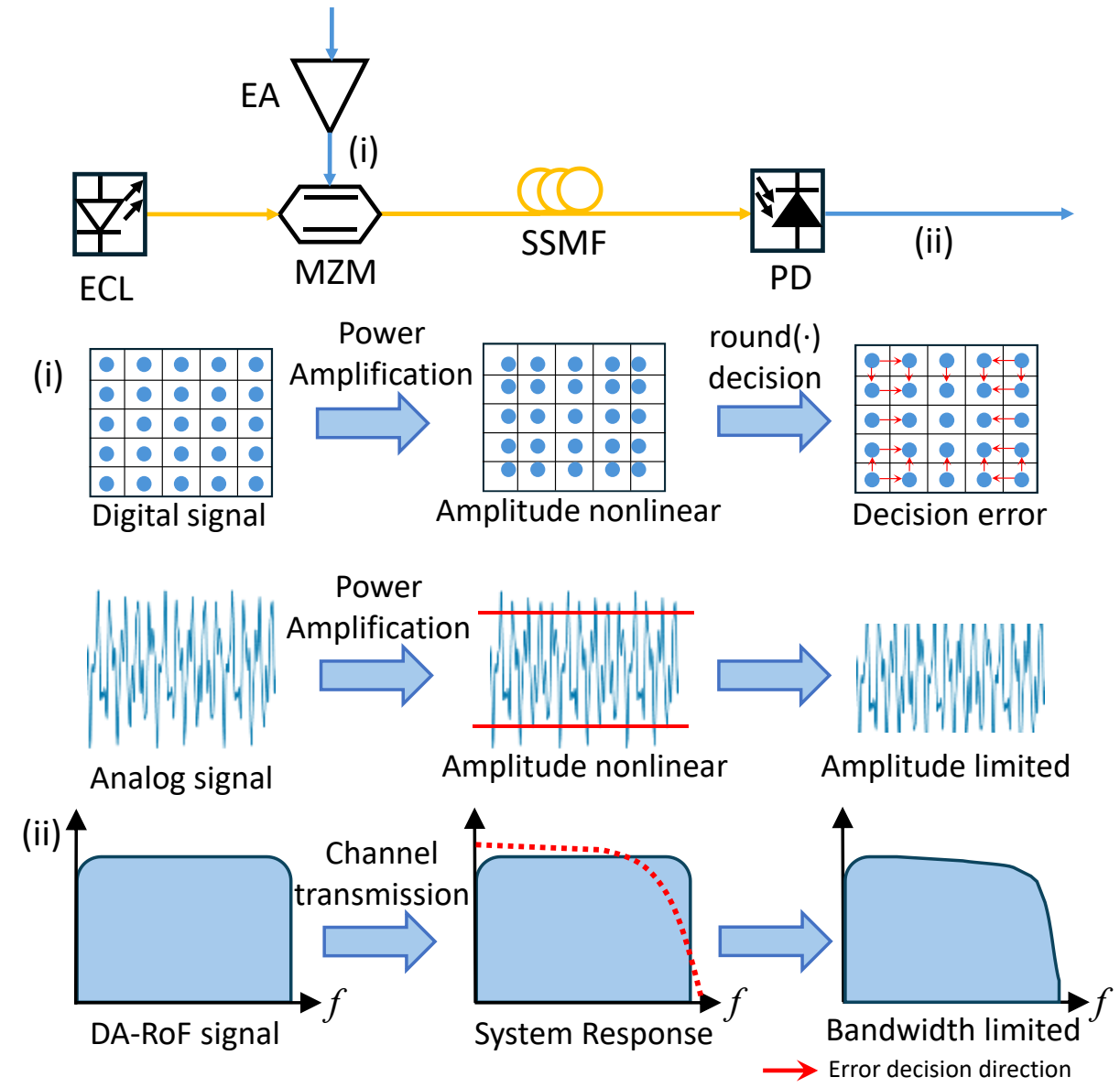

Fig. 3. System impairments in IMDD DA-RoF system: (i) amplitude nonlinear of digital and analog signals induced by amplifier; (ii) limited bandwidth induced by system components. ECL: external cavity laser; EA: electrical amplifier; MZM: Mach-Zehnder Modulator; SSMF: standard single-mode fiber; PD: photodiode.

In most cases, the decision errors occur between neighboring symbol levels, whereas cross-level errors are negligible. Thus, we assume error distance $d=1$, and the corresponding noise power caused by decision errors can be written as

$$P_{ND,i} = E[q_i^2] = SER_i \cdot (\frac{1}{rf_i^2}) \quad (6)$$

For analog part, the noise power $P_{NA}$ can be calculated as

$$P_{NA} = \frac{P_N}{\prod_{i=1}^{N} sf_i^2} \quad (7)$$

According to [37], the SNR of recovered wireless signal can be expressed as

$$SNR_w = \frac{P_s}{\sum_{i=1}^{N} P_{ND,i} + P_{NA}} = \frac{1}{\sum_{i=1}^{N} \frac{SER_i}{rf_i^2 P_N} + \frac{1}{\prod_{i=1}^{N} sf_i^2}} \cdot \frac{P_s}{P_N} \quad (8)$$

where $P_s$ is the power of original wireless signal. From formula (8), it can be observed that after transmission and demodulation, the SNR gain of recovered wireless signal is related to the RFs, SFs, and SER at a fixed $N$. RFs, which determine the digital part order, in turn affects the overall SER in the system, leading to a mutually coupled relationship. In the next section, we analyze the system impairments and discuss how these impairments influence the SER, thereby providing insights into maximizing the overall SNR.

### *B. System impairments model*

In our experiment, we implement an IMDD DA-RoF system, as shown in Fig. 3. As mentioned above, the main impairments in the system are amplitude nonlinearity and ISI.

The transmitted DA-RoF electrical signal is first amplified by an electrical amplifier (EA). Because the amplifier has a limited maximum output power, the high-amplitude portions of the signal experience amplitude clipping, which is called as amplitude nonlinear. We model the transmitter amplifier as a nonlinear AM/AM mapping $f(\cdot)$, such that the output is [38]

$$f(x)=\frac{Gx}{(1+(\frac{|x|}{A_{sat}})^{2p})^{1/2p}} \tag{9}$$

where $x$ is the signal to be amplified, $G$ is the small-signal gain, $A_{sat}$ is the saturation amplitude, and $p$ controls the softness of compression. For small $|x|$, $f(x)=Gx$; for large $|x|$, $f(x)$ flattens toward a constant magnitude. As illustrated in Fig. 3 (i), for the digital part, the outer constellation points are compressed toward the center, so that during the subsequent $round(\cdot)$ decision, these outer points may be decided as inner points. For the analog part, the high-level portions are directly clipped, causing loss of the peak information.

To mitigate this nonlinear effect, different strategies are applied to the digital and analog parts. For the digital part $S_{D,i}$, we focus on the high-level portion of signal, which we denote as $S_{HD,i}$. $S_{HD,i}$ is typically a fixed-level value for digital signal. Therefore, we introduce a GS factor $\beta$ to make $f(\beta S_{HD,i})=GS_{HD,i}$. According to formula (9), we can obtain

$$\beta=(1-(\frac{|S_{HD,i}|}{A_{sat}})^{2p})^{-1/2p} \tag{10}$$

Formula (10) indicates that, for the digital part, the nonlinearity can be mitigated by stretching the outermost constellation points. For the analog part, the amplitude of $S_A$ is controlled by $sf_N$ according to formula (2). Since analog part and the digital parts share the same time-division multiplexed input to the amplifier, $sf_N$ can be dynamically adjusted according to the amplitude of the digital parts to regulate the analog part power and thus mitigate nonlinear effects.

The ISI originates from the fact that the system bandwidth is smaller than the signal bandwidth, as illustrated in Fig. 3 (ii). When higher-rate signals are transmitted under a fixed system bandwidth, more severe ISI is inevitably introduced. Post-equalization is commonly used to mitigate ISI; however, as ISI becomes stronger, the required post-equalizer becomes increasingly complex. In fronthaul systems, we aim to minimize the cost and computational burden at the RUs side. Therefore, we introduce a pre-equalization scheme to address this issue. In this paper, we define the tap coefficients of the FIR pre-equalizer as

$$\mathbf{g}=[g_1,g_2,...,g_K] \tag{11}$$

where $K$ denotes length of pre-equalizer taps, which is usually odd.

### *C. Parameter-Dependent SNR Analysis*

To clarify the coupling relationship among the transmitter parameters, we start from the recovered SNR expression in (8). According to (8), the recovered SNR is directly related to the SER of the digital parts and the noise of the residual analog part. Since the system impairments discussed above, including transmitter nonlinearity and bandwidth-induced ISI, mainly affect the digital-part decision accuracy through SER, the coupling mechanism can be analyzed from the SER of each digital part. For the *i*-th digital part, the received symbol before hard decision can be written as

$$\tilde{d}_i[k]=d_i[k]+e_i[k] \tag{12}$$

where $d_i[k]$ is the ideal digital symbol and $e_i[k]$ denotes the equivalent decision disturbance caused by additive noise, transmitter nonlinearity, and bandwidth-induced ISI. The effective variance of $e_i[k]$ can be approximated as

$$\sigma_i^2\approx\sigma_{\mathrm{Noise},i}^2+\sigma_{\mathrm{NL},i}^2+\sigma_{\mathrm{ISI},i}^2 \tag{13}$$

Here, $\sigma_{\mathrm{NL},i}^2$ is related to the amplitude distribution and PAPR of the transmitted DA-RoF waveform, while $\sigma_{\mathrm{ISI},i}^2$ depends on the bandwidth-limited channel response and the pre-equalization coefficients. Therefore, RF, SF, GS factor, and pre-equalization jointly affect the effective decision disturbance. Assuming that the effective decision disturbance follows a Gaussian distribution and that neighboring-level decision errors dominate, the SER of the *i*-th digital part can be approximated by the standard square-QAM expression [39]:

$$SER_i\approx1-[1-2(1-\frac{1}{\sqrt{2rf_i-1}})Q(\frac{\Delta_i}{2\sigma_i})] \tag{14}$$

where $Q(\frac{\Delta_i}{2\sigma_i})$ is the Gaussian Q-function, and $\Delta_i$ is the effective decision distance after ISI and nonlinear compensation, where the ISI and nonlinear compensation are controlled by the GS factor $\beta$ and the pre-equalization coefficient $\mathbf{g}$, respectively. This expression shows that the SER is affected by both the effective decision distance and the effective disturbance variance. Specifically, RF changes the digital constellation order and decision distance, SF affects the power allocation and PAPR, GS factor modifies the effective constellation spacing under nonlinear compression, and pre-equalization affects the ISI term. Thus, the SER can be expressed as

$$SER_i=\Phi_i(rf_i,sf_i,\beta,\mathbf{g}) \tag{15}$$

Substituting this SER dependence into formula (8), the recovered SNR of wireless signal can be further expressed as a coupled function of the transmitter parameters:

$$SNR_w = \frac{1}{\sum_{i=1}^{N} \frac{\Phi_i(rf_i, sf_i, \beta, \mathbf{g})}{rf_i^2 P_N} + \frac{1}{\prod_{i=1}^{N} sf_i^2}} \cdot \frac{P_s}{P_N} = F(\mathbf{rf}, \mathbf{sf}, \beta, \mathbf{g}) \tag{16}$$

where $\mathbf{rf} = [rf_1, rf_2, ..., rf_N]$, and $\mathbf{sf} = [sf_1, sf_2, ..., sf_N]$. This non-separable relationship indicates that RFs, SFs, GS factor, and pre-equalization coefficients cannot be optimized independently, which provides the theoretical basis for formulating the transmitter optimization problem as an MDP and solving it with the proposed RL-based joint optimization framework.

## III. Problem Formulation

In this paper, we aim to design an online optimization algorithm for DA-RoF fronthaul systems to maximize the recovered wireless signal SNR at transmitter side. In particular, we make online decisions in the sense that at each transmission frame, we jointly optimize the RFs $\mathbf{rf} = [rf_1, rf_2, ..., rf_N]$, SFs $\mathbf{sf} = [sf_1, sf_2, ..., sf_N]$, GS factor $\beta$ and FIR pre-equalization coefficients $\mathbf{g}$ based on instant channel observations, without requiring knowledge of future random channel variations or system impairments.

At the beginning of frame $t$, the DA-RoF transmitter determines the online decision vector based on its observations of past and current events, including the current RFs $\mathbf{rf}_t$, SFs $\mathbf{sf}_t$, GS factor $\beta_t$ and FIR pre-equalization coefficients $\mathbf{g}_t$. The chosen decision then affects the modulation order of digital part, nonlinear distortion and ISI in future frames. Therefore, the optimal decision process naturally corresponds to an MDP, represented by a tuple $(\mathcal{S}, \mathcal{A}, \mathcal{P}, \mathcal{R}, \gamma)$. In the following, we specify the state space $\mathcal{S}$, action space $\mathcal{A}$, transition probability $\mathcal{P}$, reward function $\mathcal{R}$, and discount factor $\gamma$ for the proposed DA-RoF transmitter optimization framework.

### A. State

The system state at time $t$ is described by $s_t = (\mathbf{rf}_t, \mathbf{sf}_t, \beta_t, \mathbf{g}_t)$. The state space of the system is determined by the DA-RoF order $N$ and the taps length of pre-equalization filter $K$, and can be expressed as $\mathcal{S}_t \subseteq \mathbb{R}^{2 \cdot N + K + 1}$.

### B. Action

Based on the current state $s_t$, the outputs the variation of the state parameters as the action at time $t$. We use $\Delta$ to denote such variations, and the action can be expressed as $a_t = (\Delta\mathbf{rf}_t, \Delta\mathbf{sf}_t, \Delta\beta_t, \Delta\mathbf{g}_t)$. In our architecture, the action space is one-to-one mapped to the state space.

### C. Transition Probability

In our DA-RoF control formulation, the environment transition is deterministic and governed by the physical signal evolution and parameter update where the transition probability is implicitly captured by the replay buffer. Since the DA-RoF transmission and adaptation process is deterministic, the state transition probability is given by

$$P(s_{t+1} \mid s_t, a_t) = \delta(s_{t+1} - s_t + a_t) \tag{17}$$

### D. Reward Function

The reward function depends on the SNR of the recovered wireless signal. In compliance with mobile fronthaul requirements, the SNR is computed from the error vector magnitude (EVM) [40]. Specifically, for a block of $M$ symbols with ideal constellation points $\{s_m\}$ and detected symbols $\{\hat{s}_m\}$, the EVM is defined as

$$EVM = \sqrt{\frac{\sum_{m=1}^{M} \|s_m - \hat{s}_m\|^2}{\sum_{m=1}^{M} \|s_m\|^2}} \tag{18}$$

the SNR can be approximated from EVM by

$$SNR = -20\log_{10}(EVM) \tag{19}$$

At time $t$, the SNR of the recovered wireless signal depends only on the current state and action. According to the standards [40], different QAM orders require different SNR thresholds. Therefore, the reward function comprises two terms: (i) enforcing the instantaneous SNR at time $t$ to exceed the modulation-specific threshold, and (ii) encouraging monotonic improvement over the previous time step.

**Algorithm 1** Training process of DA-RoF fronthaul agent

1: **Initialize** evaluate Q-network $Q_{\text{eval}}(s, a; \theta)$, target Q-network $Q_{\text{tar}}(s, a; \bar{\theta})$, set $\bar{\theta} \leftarrow \theta$, replay buffer $\mathcal{D}$, and set global step counter $n \leftarrow 0$.
2: **repeat**
3: Reset DA-RoF system; obtain initial state $s_0$.
4: **for** $t = 0$ to $T_{\max} - 1$ **do**
5: Compute $\varepsilon_t = \max(\varepsilon_{\text{end}}, \varepsilon_{\text{start}} e^{-\varepsilon_{\text{decay}} t})$.
6: Select $a_t$ based on (17).
7: Apply $a_t$ to DA-RoF system.
8: Perform N-order DA-RoF modulation and fronthaul transmission.
9: Recover wireless signal and measure $SNR_{t+1}$.
10: Compute reward $r_{t+1}$ based on (15).
11: Obtain next state $s_{t+1}$ and store $(s_t, a_t, r_{t+1}, s_{t+1})$ into buffer $\mathcal{D}$.
12: **end for**
13: $n \leftarrow n + 1$.
14: **if** size($\mathcal{D}$) $\geq B$ **then**
15: Sample mini-batch $\{(s_i, a_i, r_{i+1}, s_{i+1})\}_{i=1}^{B}$ from $\mathcal{D}$.
16: Obtain Q-value for each action based on (18).
17: Compute target Q-value based on (19).
18: Compute loss based on (20).
19: Update parameters based on (21).
20: **end if**
25: **if** $n$ mod $K_{\text{tar}} = 0$ **then**
26: Update target network: $\bar{\theta} \leftarrow \theta$
27: **end if**
28: **until** the objective function in (16) converges.

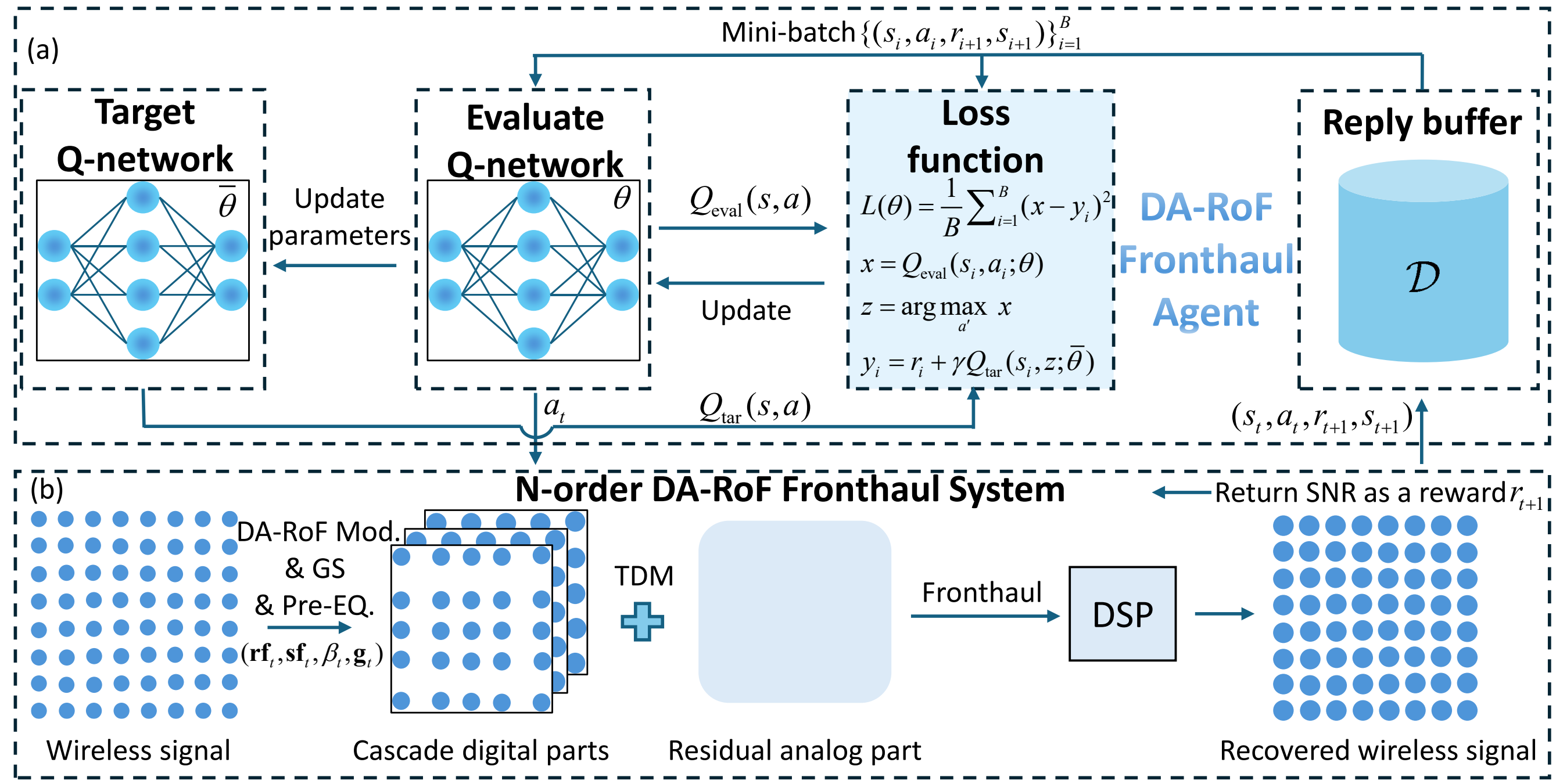

Fig. 4. Proposed RL-enabled DA-RoF fronthaul architecture. (a) DA-RoF fronthaul agent. (b) N-order DA-RoF fronthaul system.

$$r_t(s_t,a_t)=\alpha[SNR_t(s_t,a_t)-SNR_{threshold,O}] \\ +(1-\alpha)[SNR_t(s_t,a_t)-SNR_{t-1}(s_{t-1},a_{t-1})] \tag{20}$$

where $SNR_t(s_t,a_t)$ denotes the SNR of recovered wireless signal at time $t$. $SNR_{threshold,o}$ denotes the required SNR threshold of $O$-order QAM. $\alpha$ is a trade-off parameter $\alpha\in(0,1)$.

*E. Problem Formulation*

The objective of the DA-RoF fronthaul agent is to learn an optimal policy $\pi:\mathcal{S}\to\mathcal{A}$ that determines the transmitter parameters based on the current system state. Specifically, the optimal policy maximizes the expected discounted cumulative reward, formulated as

$$\underset{\pi}{\text{maximize}}\ \mathbb{E}[\textstyle\sum_{t=0}^{\infty}\gamma^t r_t(s_t,a_t)] \tag{21}$$

where $\gamma\in(0,1]$ is the discount factor that balances long-term performance and short-term reward. In other words, the agent aims to make sequential decisions that not only ensure high instantaneous recovered SNR, but also improve the overall transmission robustness over time.

The MDP-based formulation enables a reinforcement-learning-driven adaptive DA-RoF transmission strategy. Instead of relying on fixed modulation parameters or manually tuned compensation blocks, the learning agent autonomously adjusts RFs, SFs, GS factor, and pre-equalization coefficients according to instant system feedback. As a result, the proposed framework can optimize DA-RoF modulation and compensate for nonlinearity, and bandwidth-induced ISI, supporting intelligent and low-complexity transmitter optimization for next-generation RoF fronthaul systems.

## IV. The RL-Enabled DA-RoF Fronthaul Agent

In this section, we present the architecture of the RL-enabled DA-RoF fronthaul agent. We adopt a RL-enabled framework to solve the optimization problem in formula (16). We design a model-free learning agent that optimizes RFs, SFs, GS factor, pre-equalization coefficients based solely on end-to-end SNR feedback. Unlike conventional fixed DSP configurations or heuristic tuning, the proposed agent continuously interacts with the physical link and evolves toward the optimal operating point without requiring channel state information. Fig. 4 illustrates the overall workflow of the proposed RL-enabled DA-RoF fronthaul agent, which consists of two subsystems: (a) the RL agent composed of evaluate and target Q-networks, and (b) the $N$-order DA-RoF fronthaul system. The value-based DQN framework is selected according to the discrete nature of the DA-RoF parameter-control problem. In this work, each transmitter parameter is constrained within a predefined physical range and updated by a fixed step size. Therefore, each action corresponds to increasing or decreasing one parameter, or keeping all parameters unchanged, resulting in a finite discrete action space, which can convert the underlying continuous control problem into a finite-domain MDP. Compared with policy-gradient or continuous actor-critic methods, which are more suitable for continuous action control, DQN provides a simpler and more stable solution for the discrete parameter-adjustment problem considered here. Furthermore, the dueling-DQN architecture is particularly suitable for the coupled DA-RoF parameter space. In this problem, many local parameter-adjustment actions may produce similar immediate SNR changes, especially when the system is close to a high-SNR operating region. The dueling structure separately estimates the state value and the action advantage, enabling the agent to evaluate the overall quality of the current transmitter state while identifying the relative contribution of each action. This value-advantage decomposition helps improve learning efficiency and convergence stability in the discrete and coupled DA-RoF control problem. Therefore, we employ a dueling-DQN architecture, which separates state-value estimation from action-advantage evaluation and helps the agent focus on the most influential parameters by prioritizing action advantages under high-impact states, thereby accelerating convergence and improving stability in the coupled system.

Next, we introduce the entire training process for proposed DA-RoF fronthaul agent. The training follows a RL pipeline consisting of experience replay, target network update, and Q-value estimation with separate value and advantage branches. The dueling-DQN consists of an evaluate network $Q_{\text{eval}}(s,a;\theta)$ and a target network $Q_{\text{tar}}(s,a;\bar{\theta})$. Each network is divided into a shared feature extraction layer and two parallel fully connected branches. A value branch is estimated the scalar state value $V(s)$ and an advantage branch is estimated the relative importance $A(s,a)$ of each action. The two networks are initialized with random weights, and the target network is copied from the evaluate network: $\bar{\theta} \leftarrow \theta$. A replay buffer $\mathcal{D}$ capacity $C$ is also created to store state transitions $(s_t,a_t,r_{t+1},s_{t+1})$. At the beginning of each episode, the DA-RoF system is reset to obtain the initial state $s_0$. At each time step $t \in [0,T_{\max}]$, where $T_{\max}$ denotes max time step in an episode, the agent selects an action $a_t$ according to the $\varepsilon$-greedy policy:

$$a_t = \begin{cases} \text{random action}, & \text{random}() < \varepsilon_t \\ \arg\max_a Q_{\text{eval}}(s_t,a;\theta), & \textit{otherwise} \end{cases} \tag{22}$$

where $\varepsilon_t = \max(\varepsilon_{\text{end}}, \varepsilon_{\text{start}} e^{-\varepsilon_{\text{decay}} t})$ controls the exploration rate. The selected action $a_t$ updates the system parameters, the DA-RoF signal is transmitted and demodulated, and the recovered $SNR_{t+1}$ is measured. The reward $r_{t+1}$ is calculated as previously defined in formula (15). The new transition is then stored in the replay buffer $\mathcal{D}$. When an episode ends and the replay buffer contains at least $B$ samples, a mini-batch $\{(s_i,a_i,r_{i+1},s_{i+1})\}_{i=1}^{B}$ is drawn to perform a Q-value update. Each Q-value in the dueling architecture is computed from the value and advantage branches as:

$$Q(s_i,a_i;\theta) = V(s_i;\theta_V) + [A(s_i,a_i;\theta_A) - \frac{1}{|\mathcal{A}|}\sum_{a'} A(s_i,a';\theta_A)] \tag{23}$$

where $\theta_V$ and $\theta_A$ denote the parameters of the value and advantage branches, respectively. The target Q-value is obtained from the target network:

$$y_i = r_i + \gamma Q_{\text{tar}}(s_i, \arg\max_{a'} Q_{\text{eval}}(s_{i+1},a';\theta);\bar{\theta}) \tag{24}$$

The loss function for the mini-batch is the mean squared error:

$$L(\theta) = \frac{1}{B}\sum_{i=1}^{B}(Q_{\text{eval}}(s_i,a_i;\theta) - y_i)^2 \tag{25}$$

Finally, the network parameters are updated using stochastic gradient descent:

$$\theta \leftarrow \theta - \mu\nabla_\theta L(\theta) \tag{26}$$

To stabilize learning, the target network is periodically synchronized every $K_{\text{tar}}$ steps: $\bar{\theta} \leftarrow \theta$. This soft-update strategy prevents oscillation and ensures smoother convergence. The training process of the proposed DA-RoF fronthaul agent is summarized in Algorithm 1.

## V. Simulation Investigations

To further justify the selection of the RL algorithm and evaluate its advantage over other optimization methods, a simulation-based benchmark was carried out using VPI TransmissionMaker. The simulated link follows the same DA-RoF fronthaul configuration considered in the experimental study. In the algorithm comparison, the initial transmitter parameters are set to $(\mathbf{rf}_0, \mathbf{sf}_0, \beta_0, \mathbf{g}) = (3.0, 4.0, 1.0, [0,0,0,1,0,0,0])$, corresponding to an initial SNR of approximately 27.48 dB, to avoid starting from an overly favorable operating point. Three RL algorithms, including vanilla DQN, double DQN, and dueling DQN, were evaluated. Each algorithm was trained for 200 episodes, with

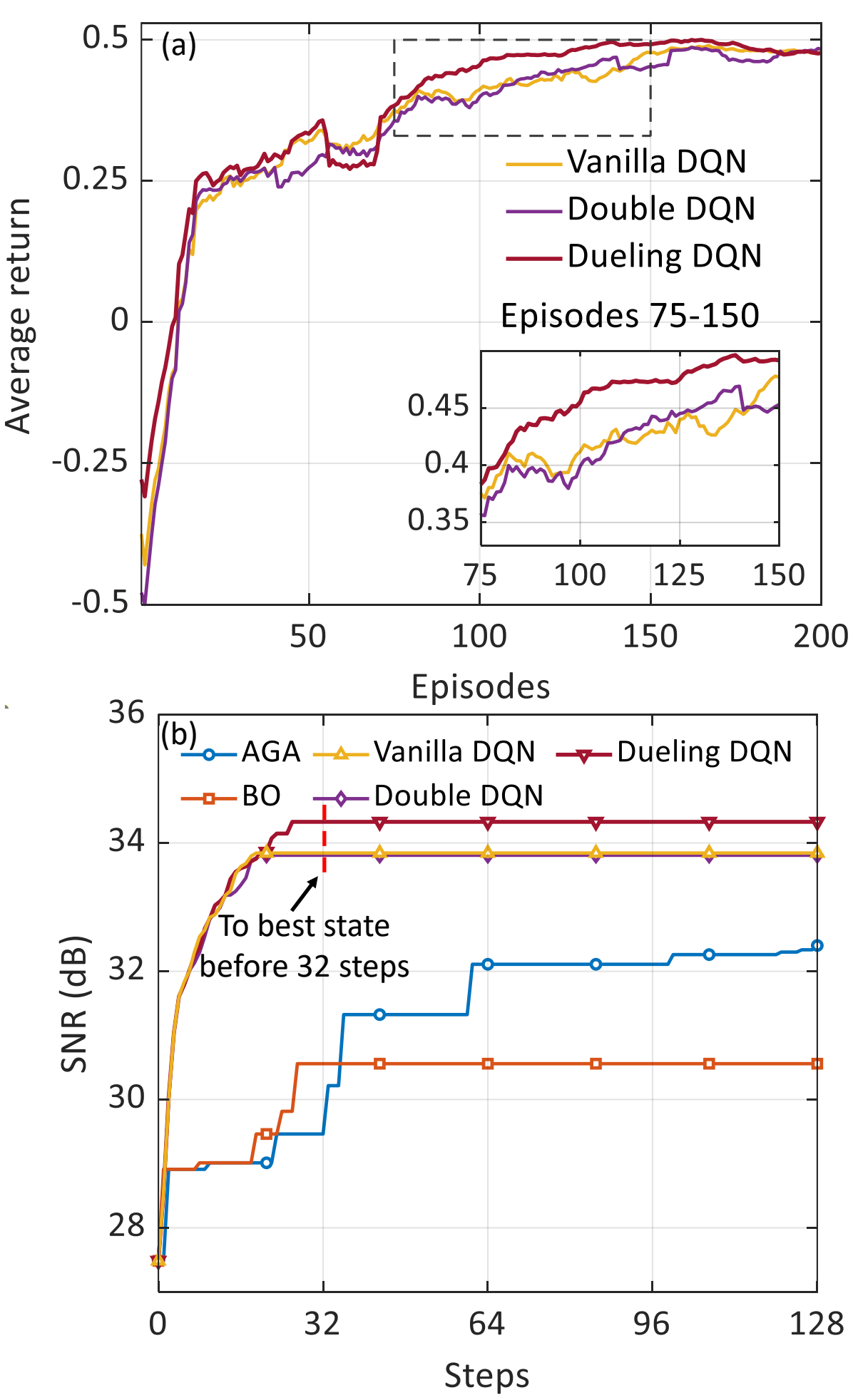

Fig. 5. Simulation comparison of optimization algorithms. (a) Average return versus training episode for different RL algorithm. (b) Best SNR versus interaction steps for AGA, BO, Vanilla DQN, Double DQN, and Dueling-DQN.

32 interaction steps in each episode. As shown in Fig. 5(a), dueling DQN achieves a higher average return in the early training stage and converges faster than the other two DQN variants, indicating better learning efficiency in the coupled DA-RoF parameter space.

After training, we further compared the inference performance of the three RL algorithms with two conventional search-based optimization methods, namely adaptive genetic algorithm (AGA) and Bayesian optimization (BO), as shown in Fig. 5(b). Within the 128 interaction steps, AGA and BO only achieve SNRs of 34.18 dB and 31.42 dB, respectively, while the trained dueling-DQN agent reaches the high-SNR region before 32 interaction steps and maintains a stable optimization trajectory. More importantly, AGA and BO only search for the optimal parameter set under the current condition and do not learn a reusable state-dependent parameter-update policy. Once the link condition or initial state changes, the search process needs to be repeated. This makes the proposed dueling-DQN agent more suitable for fast deployment and adaptive parameter optimization in practical DA-RoF fronthaul systems. Moreover, the value-advantage decomposition of dueling DQN is well suited to the discrete parameter-adjustment problem considered here, where many local actions may lead to similar immediate SNR changes but differ in their long-term contribution. Therefore, dueling DQN is selected as the main RL optimization algorithm in this work.

## VI. Experimental Evaluation

This section aims to experimentally validate the effectiveness of the proposed DA-RoF fronthaul agent on an IMDD optical communication system demo platform. We provide a detailed analysis and comparison of the DA-RoF fronthaul agent based on the experimental results.

### *A. Experimental setup and DSP processing*

Fig. 6(d) illustrates the experimental setup of the IMDD system. At the transmitter, the electrical signal is first generated and output by a 65-GSa/s digital-to-analog converter (DAC), followed by amplification through an EA. A thin-film lithium-niobate Mach–Zehnder modulator (TFLN-MZM) with a 3-dB bandwidth of 25 GHz is employed to modulate the optical carrier emitted from an external-cavity laser (ECL) centered at 1550 nm. The modulated optical signal is then boosted by an erbium-doped fiber amplifier (EDFA) before being transmitted over a 1-km standard single-mode fiber (SSMF) link. At the receiver, a variable optical attenuator (VOA) is used to control the received optical power (ROP). The optical signal is converted to the electrical domain by a 60-GHz photodiode (PD) and subsequently sampled by a 100-GSa/s analog-to-digital converter (ADC), whose bandwidth is 12.5 GHz, for offline digital signal processing (DSP). Fig. 6 (a) and (c) show the DSPs of DA-RoF system in the transmitter and receiver. In Tx-side DSP, a pseudorandom bitstream (PRBS) data is generated. The bits are first mapped into 1024-, 4096-, 16384- and 65536-QAM symbols corresponding to 1-, 2-, 3- and 4-order DA-RoF signal. The OFDM modulation is employed to simulate the aggerated wireless signals. After serial to parallel conversion, 1024-point FFT is used for OFDM modulation, in which 1000 subcarriers are loaded with symbols. Then, the OFDM signal is modulated by DA-RoF modulation according to formula (1) and (2), whose RFs and SFs are $\mathbf{rf}$ and $\mathbf{sf}$ . The digital parts of DA-RoF signal are shaped by GS factor $\beta$ . After up-sampling and interleaving the I and Q paths to form a real signal [41], the signal is pre-equalized according to taps coefficients $\mathbf{g}$ .These parameters of transmitter are guided by DA-RoF fronthaul agent. To evaluate the performance and robustness of the agent under varying signal bandwidth conditions, we conduct tests using DA-RoF signals with baudrate of 26 Gbaud, according to the system bandwidth limit. The transmitted signal is resampled to 65 GSa/s and sent to DAC. In Rx-side offline DSP, the captured signal is resampled to 2 samples per symbol. Synchronization is achieved by identifying the peak of the cross-correlation between the received signals and the synchronization sequence. The feed-forward equalizer (FFE) is employed to remove residual ISI. The equalized signal is reassembled into a complex signal, and then DA-RoF demodulation is performed according to formula (3).

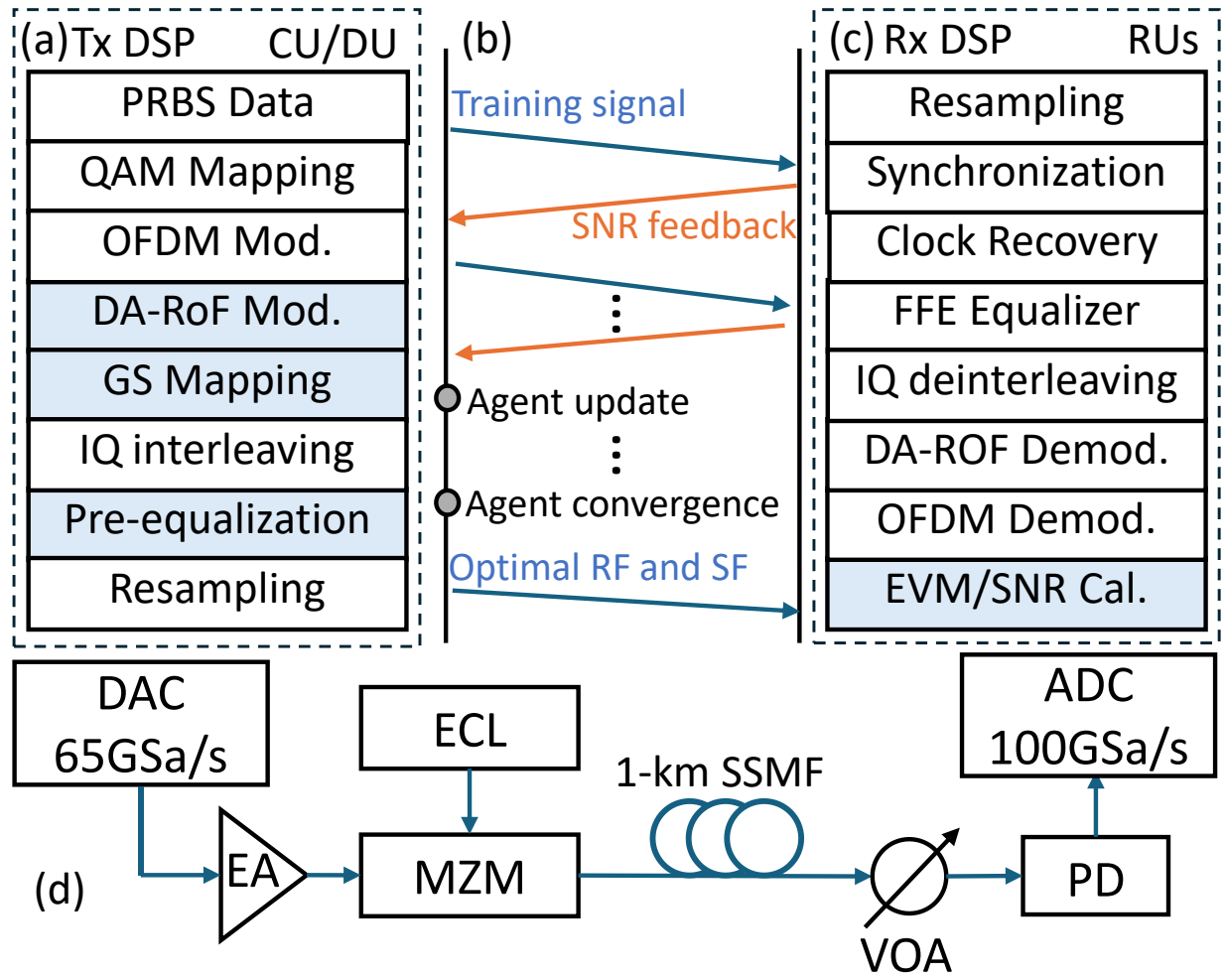

Fig. 6. (a) Transmitter (Tx) DSP at CU/DU. (b) The interaction process in training. (c) Receiver (Rx) DSP at RUs. (d) IMDD system with 1-km SSMF.

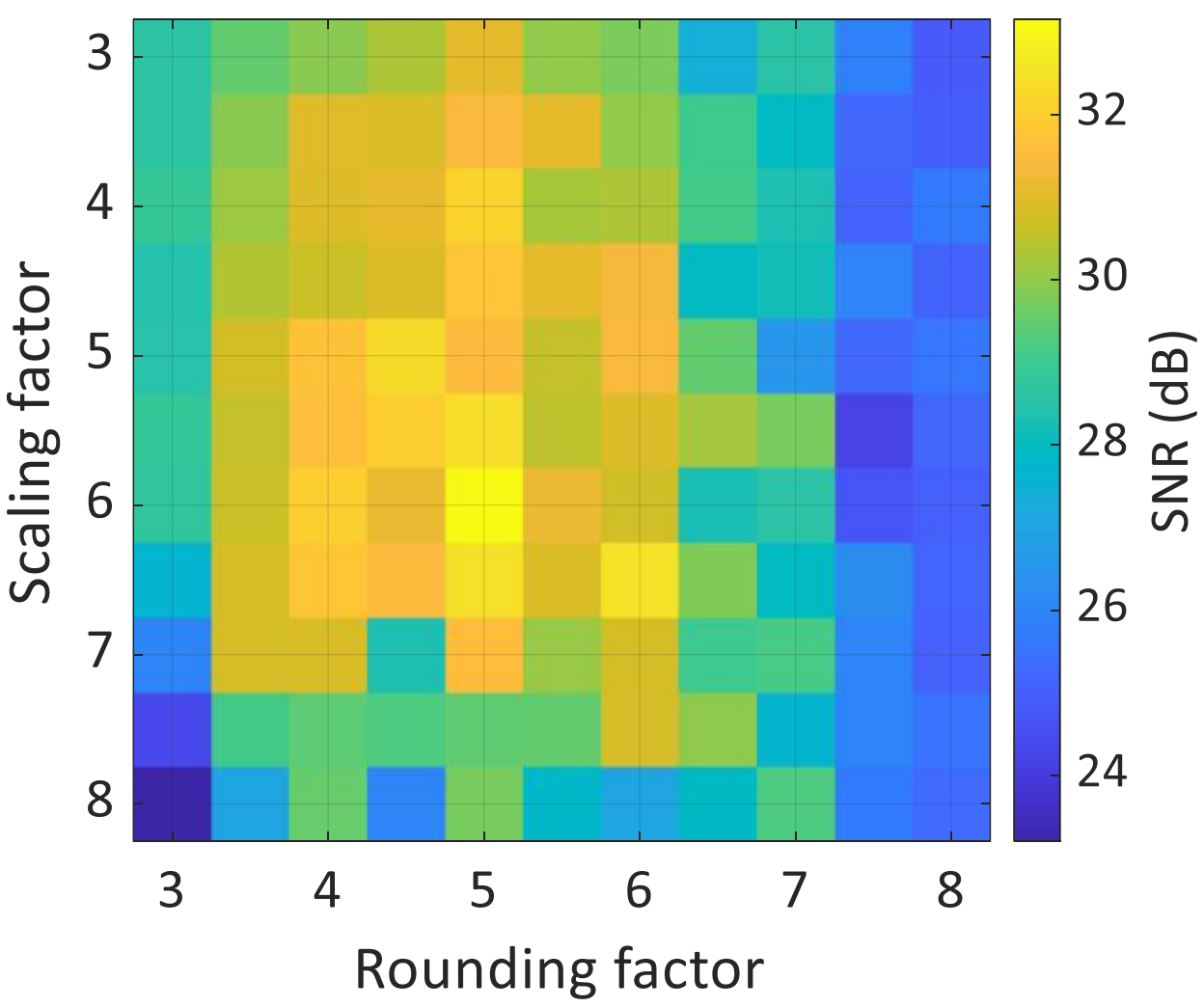

Fig. 7. The recovered SNR for the wireless signal versus different rounding and scaling factors.

Finally, the SNR of the recovered signal is calculated according to formulas (13) and (14) after OFDM demodulation.

The hardware-friendly property of the proposed framework mainly comes from the transmitter-side deployment of the RL agent and the separation between the training and inference stages. During training, the agent is executed at the CU/DU side and interacts with the optical fronthaul link through SNR feedback from the RU. During both training and inference, no additional device or DSP module is required at the RU side, and the RU only needs to feed back the measured SNR. The additional computation is centralized at the CU/DU side. The dueling-DQN is lightweight, consisting of a 13-dimensional input layer, two 128-neuron shared fully connected layers, an advantage stream with 27 outputs, and a value stream with one output. The total number of trainable parameters is 54,940. Therefore, the proposed framework introduces only limited CU/DU-side digital computation while keeping the RU-side hardware complexity unchanged, which supports its practical deployment in DA-RoF fronthaul systems.

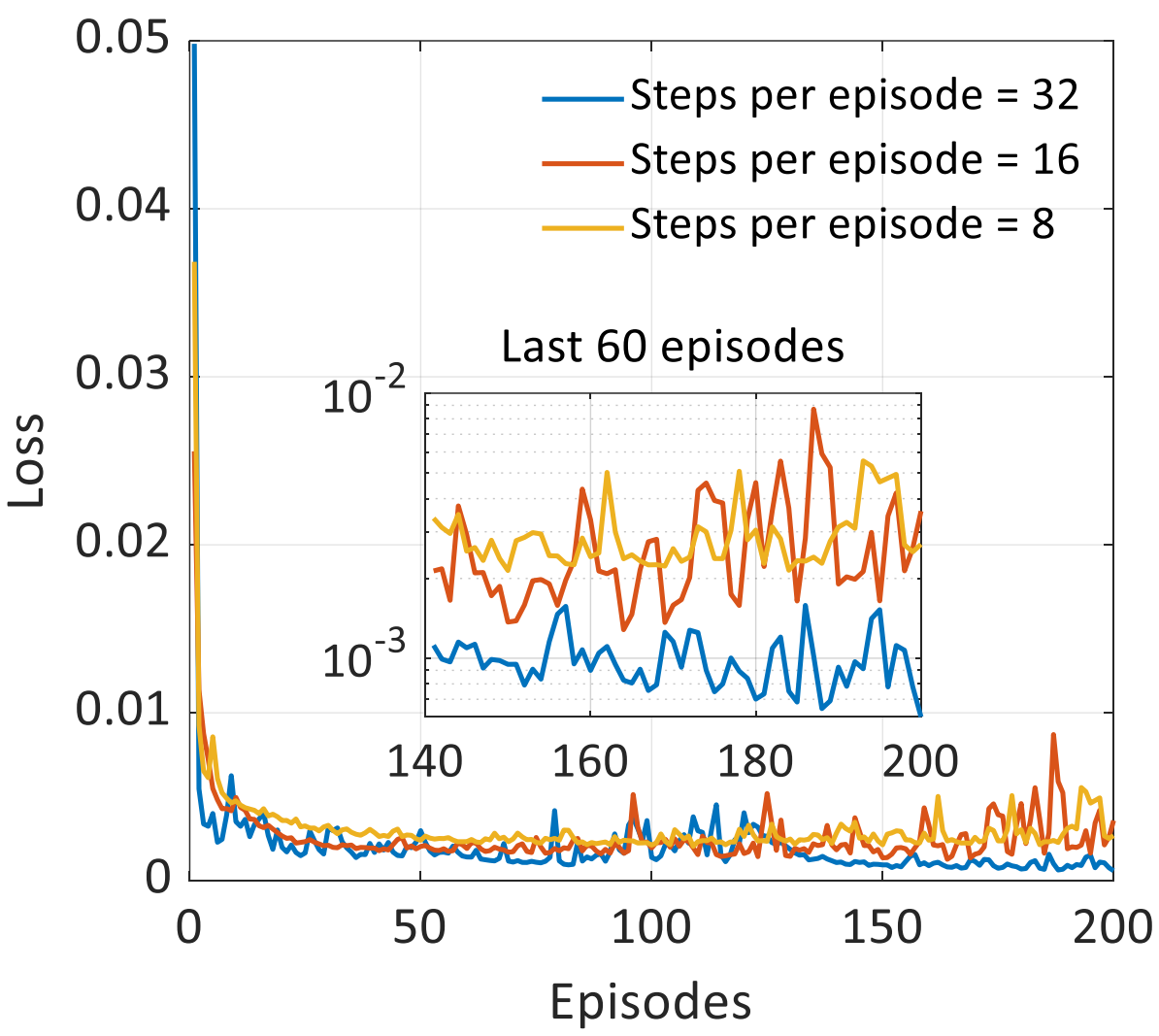

Fig. 8. Training loss of the DA-RoF fronthaul agent under different maximum step numbers per episode (8, 16, and 32).

### *B. Training of the DA-RoF Fronthaul Agent*

In this section, we present the training process and results of the agent in the experimental DA-RoF system, taking the 1-order DA-RoF case as an example. As shown in Fig. 5 (b), during the agent training phase, a training signal corresponding to the current state $s_t = (\mathbf{rf}_t, \mathbf{sf}_t, \beta_t, \mathbf{g}_t)$ is transmitted from the CU/DU to the RUs. After Rx DSP, the recovered SNR is fed back to the CU/DU as the reward. Through iterative interactions between action selection and reward feedback, the agent parameters are continuously updated until convergence is achieved. Once training is completed, the optimized agent delivers the optimal RF and SF values to the RUs for DA-RoF demodulation. At this point, the DA-RoF parameter optimization process is finalized and normal communication service can be initiated. In the results, we first disable GS and pre-equalization, and perform a grid search over the RF and SF to obtain the performance baseline of the 1-order DA-RoF signal. As shown in Fig. 7, the optimal SNR is achieved at $(\mathbf{rf}, \mathbf{sf}) = (5.0, 6.0)$, where the recovered wireless signal reaches an SNR of 33.1 dB. Based on this baseline, we use a suboptimal state $(\mathbf{rf}_0, \mathbf{sf}_0) = (3.0, 4.0)$ as the initial state to better demonstrate the effectiveness of reinforcement learning. For the GS factor, the initial value $\beta_0 = 1.0$. For the pre-equalization filter, we adopt a 7-tap FIR structure initialized as $\mathbf{g}_0 = (0,0,0,1,0,0,0)$. In summary, the initialized state is $s_0 = (\mathbf{rf}_0, \mathbf{sf}_0, \beta_0, \mathbf{g}_0)$. The trade-off parameter in the reward function is set to $\alpha = 0.8$ for all experiments. Each transmitter parameter is discretized within a physically meaningful range and updated with a fixed step size. Specifically, the $(\mathbf{rf}, \mathbf{sf})$ are both constrained within [2.0, 8.0] with a step size of 0.1, while the GS factor $\beta$ is constrained within [1.0, 1.3] with a step size of 0.025. For the 7-tap pre-equalization filter, t aps 1 and 7 are constrained within [-0.2, 0.2], taps 2 and 6 within [-0.3,0.3], taps 3 and 4 within [-0.5, 0.5], and tap 5 within [0.5, 1.0], all with a step size of 0.025. At each decision step, the agent selects one action to increase or decrease one parameter by its corresponding step size, or selects the no-operation action. Since the state contains 13 transmitter parameters in this implementation, the total number of discrete actions is 27. As shown in Fig. (7), we first monitor the evolution of the training loss to assess the stability of Q-network fitting and the convergence speed of the value function. The comparison reveals that a longer steps of an episode allows the agent to interact more extensively with the environment, enabling more accurate estimation of action values and significantly reducing the training loss. Next, we compute the average return of each episode, as shown in Fig. (8), to evaluate the overall performance of the learned policy across the parameter space. The experimental results show that the average return gradually increases as training progresses, indicating that the agent successfully transitions from a random policy to a high-performance one. The training process is considered converged using a sliding-window criterion. Specifically, a window length of 30 episodes is used to calculate the average episodic return. When the averaged return fluctuation over 10 consecutive windows is less than 0.05, the training is regarded as converged. Furthermore, we examine the influence of steps per episode on policy quality and convergence stability. Agents trained with shorter steps per episode update their policies more frequently in the early stage, but their limited exploration leads to inferior final returns and less stable loss curves. In contrast, longer steps per episode provide richer state visitation, enabling the agent to achieve a higher performance ceiling and lower fluctuation in the later stage. This illustrates that, for a DA-RoF system with coupled nonlinearities and multidimensional parameter interactions, increasing the depth of agent-environment interaction is crucial for learning an optimal control policy. Overall, the training process can converge within 6400 interactions with the RU, corresponding to 32 steps × 200 episodes. After convergence, the inference process can reach the optimized state within 32 interaction steps. In contrast, exhaustive joint parameter search requires testing all possible combinations of RFs, SFs, GS factor, and pre-equalization coefficients. Even if the pre-equalization taps are optimized separately by gradient descent, the remaining parameters, including RFs, SFs, and the GS factor, still require exhaustive interaction-based search. For example, in the 1-order DA-RoF case, RF and SF each have 61 candidate values, and the GS factor has 13 candidate values, leading to 61×61×13=48,373 required online SNR measurements. This is still much larger than the RL training budget of 6400 interactions. Moreover, such a two-stage optimization strategy separates pre-equalization from RF/SF/GS optimization and therefore cannot guarantee joint optimization over the full transmitter-parameter space.

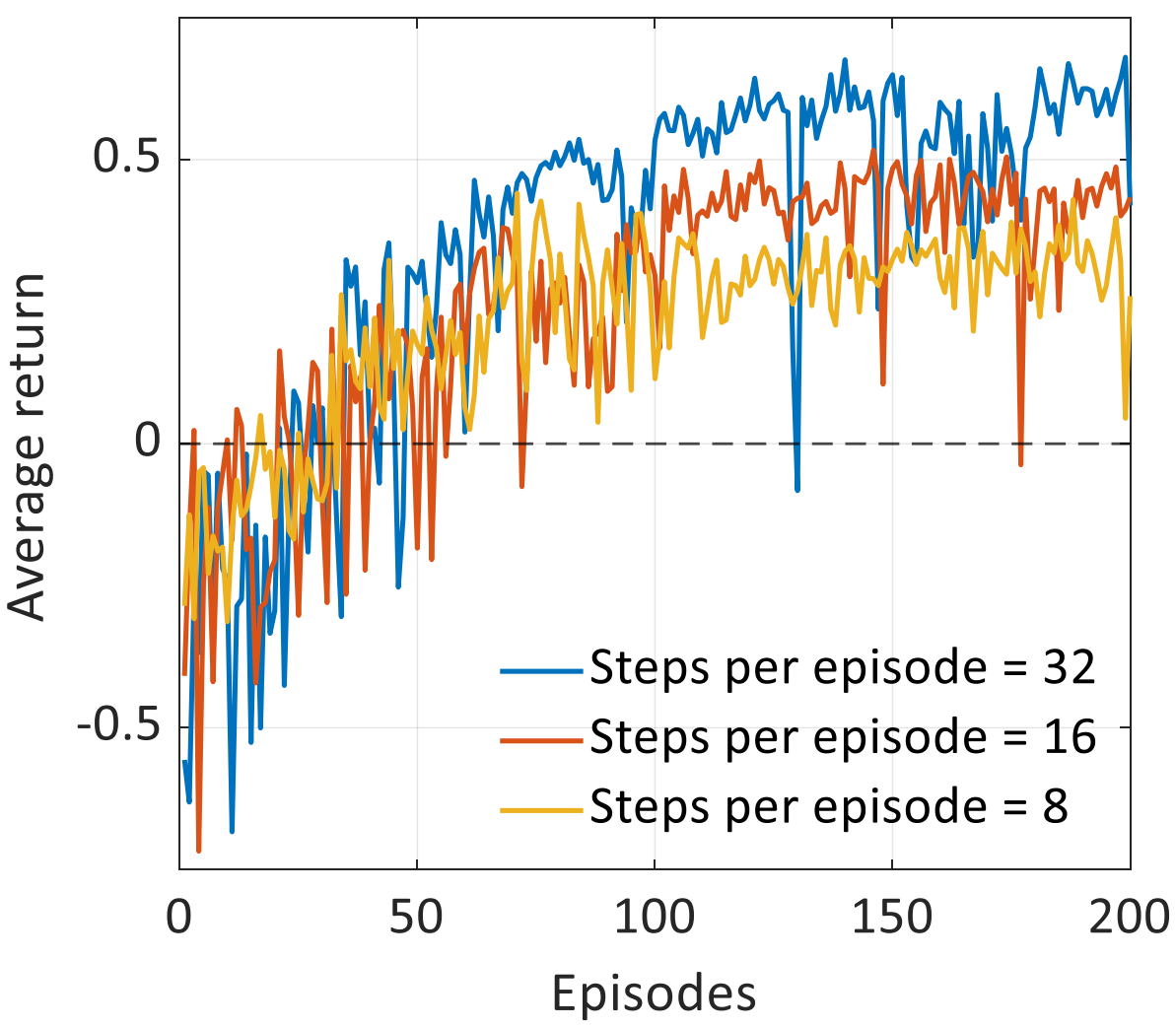

Fig. 9. Average episodic return of the DA-RoF fronthaul agent in training under different maximum step numbers per episode (8, 16, and 32).

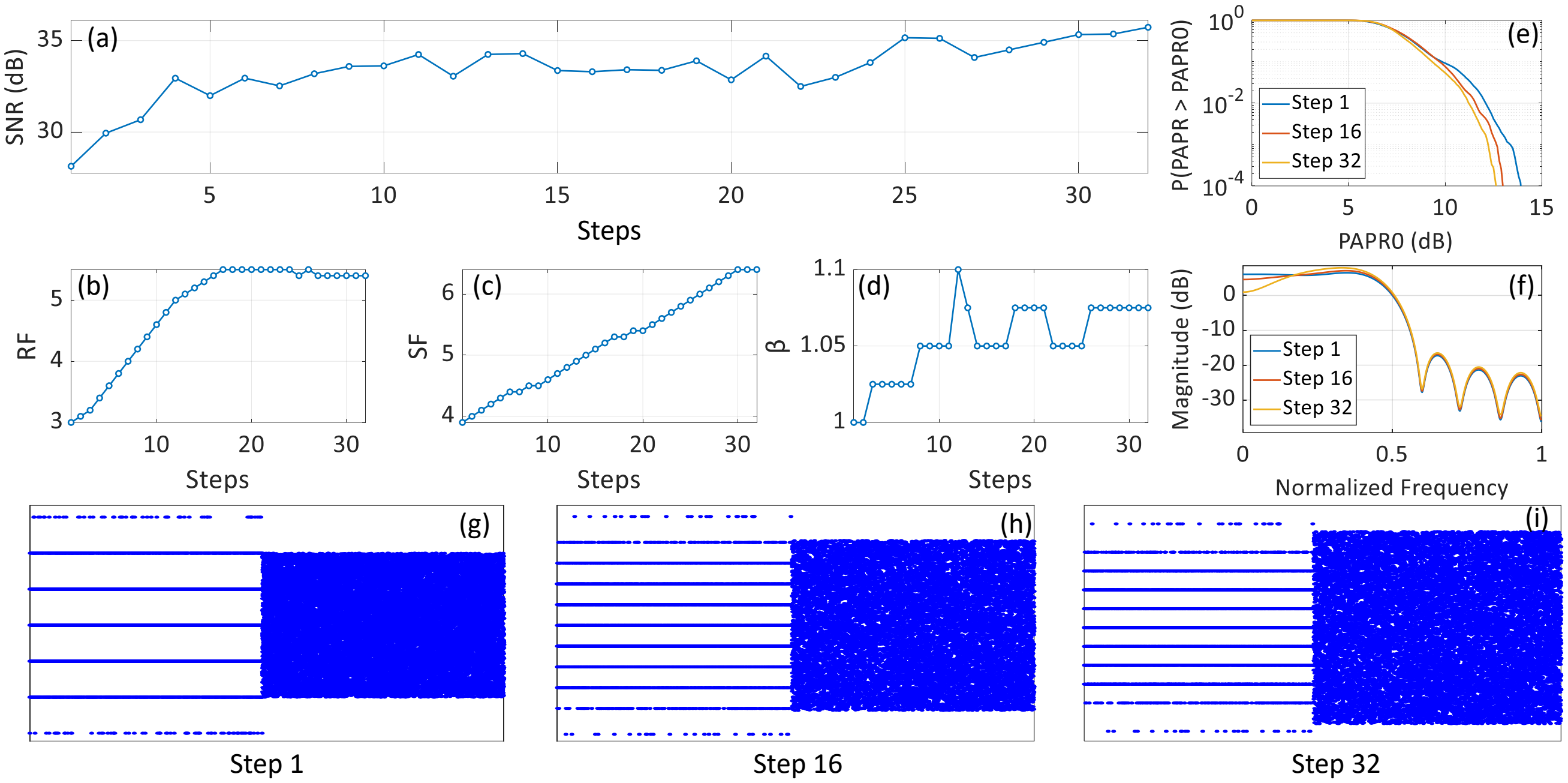

Fig. 10. (a) Evolution of the recovered SNR over decision steps; (b) Iterative update of RF; (c) iterative update of SF; (d) iterative update of shaping factor β; (e) PAPR complementary cumulative distribution function (CCDF) of transmitted DA-RoF signal at step 1, step 16 and step 32; (f) frequency response of the learned pre-equalization filter at step 1, step 16 and step 32 ; (g), (h) and (i) transmitted DA-RoF signal at step 1, step 16 and step 32.

It is worth noting that the current agent is trained and validated under the IMDD DA-RoF experimental setup used in this work. When the channel condition, hardware response, link bandwidth, or operating point changes significantly, the agent may require further fine-tuning or retraining based on new SNR feedback. Since the proposed framework relies only on measured SNR feedback rather than an explicit channel model, it is expected to be adaptable to different link conditions through additional online interaction or transfer training.

### C. Inference of the DA-RoF Fronthaul Agent

After completing the training stage, the learned policy is deployed for inference to evaluate its online control capability in the DA-RoF fronthaul system. During inference, the agent no longer performs exploration and selects actions purely based on the greedy policy $a_t = \arg\max_a Q(s_t, a; \theta^*)$, where $\theta^*$ denotes the converged parameters of the dueling DQN. Starting from an initial state, the trained agent sequentially adjusts RF, SF, GS factor, and pre-equalizer taps according to the learned state-action mappings. For each interaction, the environment provides the updated SNR, and the agent iteratively drives the DA-RoF system toward the optimal operating region without any grid search, manual tuning, or prior knowledge of the channel conditions. In Fig. 10 (a), the SNR increases steadily from approximately 29 dB to 35.8 dB as the agent iteratively updates the DA-RoF control parameters, confirming that the learned policy can autonomously drive the system toward a high-performance operating region. Fig. 10 (b)-(d) show the independent evolution of the three optimized parameters, where the RF gradually converges to around 5.4, the SF increases linearly towards 6.5, and the GS factor $\beta$ slightly adjusts around 1.075 to compensate for amplitude nonlinearity. To evaluate the temporal evolution of waveform characteristics, Fig. 10 (e) plots the complementary cumulative distribution function (CCDF) of the PAPR. The PAPR distribution progressively shifts leftward with decision steps, reflecting reduced amplitude peaks and improved robustness against nonlinear distortion. Fig. 10 (f) shows the frequency response of the learned pre-equalization filter, where the pre-equalization filter gradually converges toward a response that compensates for the high-frequency bandwidth limitation. Finally, the evolution of the transmitted DA-RoF signal in Figs. 10 (g)-(i) further validates the agent's adaptive parameter adjustment strategy. At step 1 (Fig. 10 (g)), the initial RF value is relatively small, resulting in insufficient digital resolution for the DA-RoF representation and causing noticeable symbol clustering and quantization distortion. As the decision process progresses, the agent gradually increases RF, thereby expanding the available constellation levels and reducing digital quantization loss, which leads to a higher SNR gains. Meanwhile, increasing RF inherently elevates the dynamic range of the digital portion and may increase PAPR if not properly compensated. To avoid excessive peak energy and associated nonlinear distortion, the agent simultaneously increases the SF value, ensuring proper amplitude scaling of the analog residual component. This coordinated evolution between RF, SF and GS factor results in a balanced signal structure that satisfies both SNR enhancement and PAPR constraints, ultimately producing a more compact and noise-robust distribution at step 32.

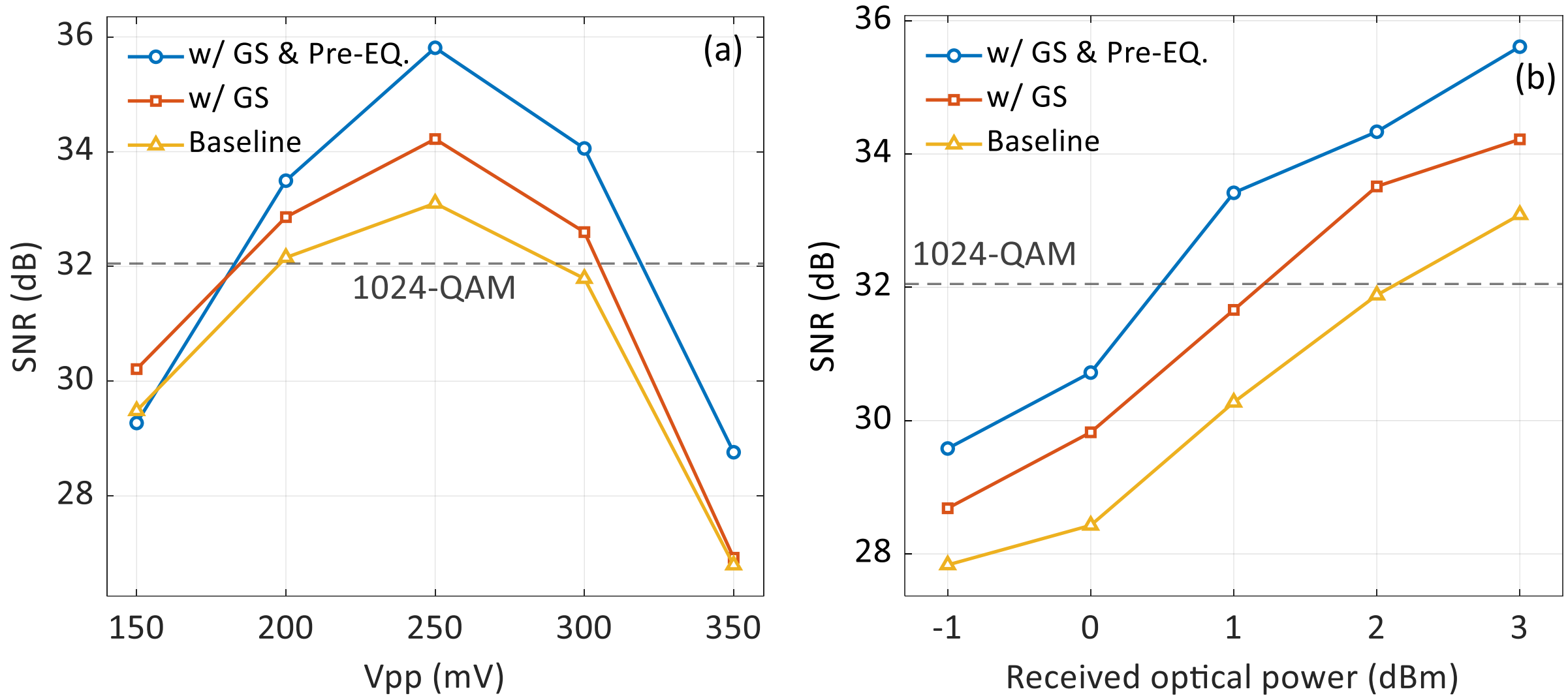

Fig. 11. (a) SNR performance of the 1-order DA-RoF signal with GS & Pre-EQ, GS only, and baseline under different Vpp values; (b) SNR performance of the 1-order DA-RoF signal with GS & Pre-EQ, GS only, and baseline versus received optical power.

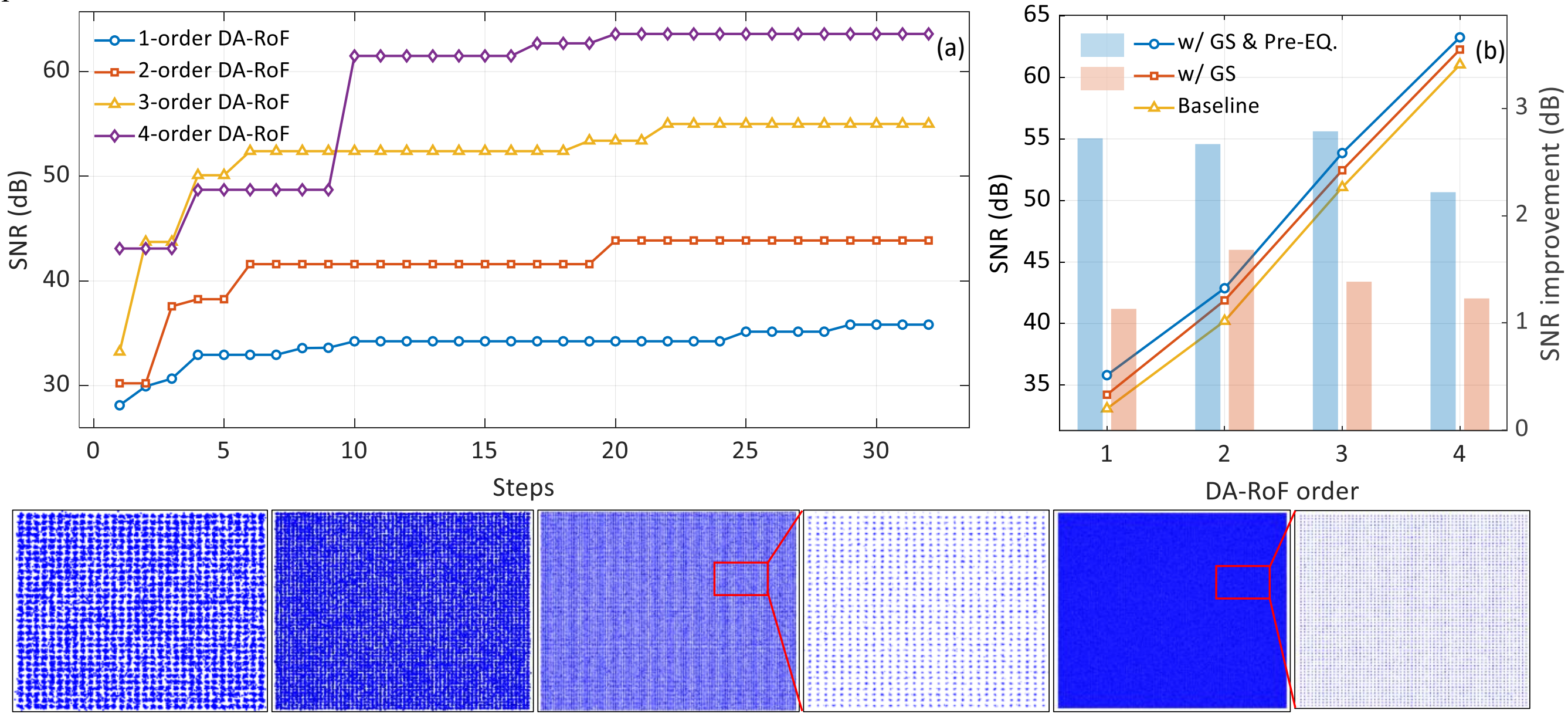

Fig. 12. (a) Best SNR evolution curves during the inference process for different DA-RoF modulation orders; (b) SNR performance with GS & Pre-EQ, GS only, and baseline versus different DA-RoF orders. Inset (i)-(vi) the constellations of recovered wireless signal (i) 1024-QAM (1-order), (ii) 4096-QAM (2-order), (iii) 16384-QAM (3-order), and (v) 65536-QAM (4-order).

The results verify that the learned policy can generalize to real transmission environments and effectively optimize the DA-RoF parameters in an online and adaptive manner.

### *D. Performance evaluation*

To evaluate the generalization capability of the agent-guided parameter optimization strategy under varying transmission conditions, we further measure the performance of the 1-order DA-RoF signal with respect to different electrical driving amplitudes peak-to-peak voltage (Vpp) and received optical powers (ROP). Three configurations are compared in this experiment: (1) GS & pre-equalization (Pre-EQ.), (2) GS only, and (3) baseline without GS and pre-equalization. The goal is to determine whether the optimal parameters recommended by the trained agent remain robust when the system operating point deviates from the training conditions. Fig. 10 (a) presents the SNR performance under different Vpp levels. All three configurations exhibit a convex relationship with Vpp, where SNR initially increases and then decreases once strong nonlinear distortion becomes dominant. The proposed GS and pre-equalization consistently achieves the highest SNR across the entire operating range, and maintains a significant margin over the 1024-QAM SNR threshold. In comparison, the GS-only configuration also provides noticeable performance improvement relative to the baseline, confirming that constellation shaping effectively alleviates nonlinear degradation. Fig. 10 (b) shows the SNR performance versus ROP. Under this condition, the GS & pre-equalization configuration continues to deliver the best performance, achieving a peak SNR of approximately 35.8 dB, followed by GS-only, and the baseline as the lowest. These results verify that the agent-recommended parameter set exhibits strong adaptability and generalization across different operating regimes. Compared with the RF/SF-only grid-search baseline, the enlarged optimization space enables the agent to simultaneously balance quantization noise, residual analog power, PAPR-induced nonlinear distortion, and bandwidth-induced ISI. The combined use of GS and pre-equalization provides complementary benefits. GS predominantly mitigates amplitude nonlinear, while pre-equalization compensates for high-frequency bandwidth limit, which both result in higher achievable SNR.

To further investigate the scalability and DA-RoF order adaptability of the proposed RL-enabled DA-RoF fronthaul optimization framework, we extend the agent from the 1-order DA-RoF modulation to higher-order cases, including 2-order, 3-order, and 4-order DA-RoF. For each DA-RoF order, the agent was trained independently based on its corresponding state-action space and signal characteristics, ensuring that the learned policy was tailored to the specific digital-analog decomposition structure.

Fig. 12 (a) shows the best SNR evolution curves during the inference process for different DA-RoF modulation orders. All modulation orders exhibit a rapid SNR increase within the initial steps, demonstrating that the agent can effectively adjust the model parameters. As the modulation order increases, the final converged SNR also increases, achieving 35.8 dB, 42.9 dB, 53.8 dB, and 63.2 dB for 1- to 4-order DA-RoF, respectively. The SNR performance for the three configurations across different modulation orders are summarized in Fig. 12 (b). Clear improvements can be observed as the modulation order increases, and the combination of GS & pre-equalization consistently provides the highest gain across all cases. The constellations of recovered wireless signal are shown in Fig. 12 (i)–(vi). For higher-order DA-RoF, the SNR of recovered wireless signal become significantly higher, progressing from 1024-QAM (1-order) to 4096-QAM (2-order), 16384-QAM (3-order), and 65536-QAM (4-order).

The proposed RL-enabled agent is evaluated on an IMDD DA-RoF optical fronthaul link, considering practical optical-link impairments such as transmitter nonlinearity, bandwidth-induced ISI, received-power variation, and different DA-RoF orders. The results demonstrate the robustness and adaptability of the proposed transmitter optimization framework under different optical-link operating conditions. The extension to multi-RU concurrent transmission and interference-aware optimization requires a more complex network-level fronthaul testbed and will be investigated in future work. Overall, these experiments confirm that the proposed RL-enabled DA-RoF fronthaul agent preserves scalability, converges rapidly across different modulation orders, and effectively unleashes the potential of high-order DA-RoF transmission. This validates the feasibility of integrating intelligent control into ultra-high-order DA-RoF systems, paving the way for future ultra-dense hybrid modulation, beyond-Terabit optical fronthaul, and hardware-adaptive next-generation access networks.

## VII. Conclusion

In this paper, we present an intelligent optimization framework for DA-RoF fronthaul systems using a dueling-DQN reinforcement learning agent. By transforming transmitter parameter configuration into a MDP, the agent autonomously learns optimal combinations of RF, SF, GS factor, and pre-equalization taps coefficients from instant SNR feedback. Experimental results validate that the proposed method substantially improves SNR compared with baseline, achieving up to 2.7 dB gain for 1- to 4-order DA-RoF. Additionally, the framework requires no channel modeling, exhaustive search, nor additional DSP complexity at the receiver, demonstrating its practicality for cost-sensitive fronthaul deployments. With its scalability and rapid convergence, the proposed approach holds strong potential for future extensions into high-order modulation format, MIMO systems, integrated photonic front-ends, and hardware-adaptive next-generation access networks.

## Acknowledgment

This work was supported in part by Mobile Information Networks-National Science and Technology Major Project(2026ZD1308000), Natural Science Foundation of Shanghai (24ZR1490500), the Major Key Project PCL, and AI for Science Program, Shanghai Municipal Commission of Economy and Informatization (2025-GZL-RGZN-BTBX-02025). We also thank Shanghai Innovation Institute.